\documentstyle[sprocl]{article}

\input{psfig}                

\newcommand{\BEQ}{\begin{equation}}    
\newcommand{\BEA}{\begin{eqnarray}}
\newcommand{\EEQ}{\end{equation}}      
\newcommand{\EEA}{\end{eqnarray}}
\newcommand{\eps}{\epsilon}                      
\newcommand{\rar}{\rightarrow}                   
\newcommand{\wit}[1]{\widetilde{#1}}             
\newcommand{\ket}[1]{\left|#1\right\rangle}      
\newcommand{\bra}[1]{\left\langle #1\right|}     

\begin{document}

\title{DOMAIN WALLS IN THE SPIN-$S$ QUANTUM ISING CHAIN
\footnote{Invited talk given at the Statphys-19 Satellite meeting on 
`Statistical Models, Yang-Baxter equations and related topics', 
8-10 Aug. 1995, Nankai University, Tianjin, China}
}

\author{Malte Henkel}

\address{Laboratoire de Physique du Solide \footnote{Unit\'e de recherche
associ\'ee au CRNS no 155}, Universit\'e Henri Poincar\'e Nancy I,
B.P. 239, F -- 54506 Vand{\oe}uvre l\`es Nancy Cedex, France}


\maketitle\abstracts{
Domain walls in the spin $S$ quantum Ising chain at zero temperature
are considered. 
Quantum effects prevent the walls from being sharp and delocalize the
exact eigenstates. The absence of a characteristic length of the
domain walls is reflected in the finite-size scaling of the spectrum
of the Hamiltonian. 
}

\section{Introduction}

Recently, there has been a growing interest in the study of interfaces 
with nontrivial geometry. Such interfaces may arise for example in domain
walls in random magnets, fluid invasion in porous media, spreading on 
heterogeneous surfaces, biological membranes and vesicles or epitaxial 
growth.\cite{exa} 
When do quantum effects play a significant role in such problems ? For 
static properties in non-random systems such as an spin $S$ antiferromagnet
with nearest neighbor exchange interaction $J$, the following regimes exist.
Near the critical temperature $T_N \sim J S^2$, thermal fluctuations dominate.
In the ordered phase, for temperatures $T_c/S \ll T \ll T_c$, quantum effects
due to the finiteness of $S$ are unimportant. For $T< T_c /S$, however, one
is in the quantum regime, where the quantum statistics of spin waves and their
interactions lead to dependences on $S$ and $T$ not present in the classical
$(S\rar\infty)$ limit. 

Useful insight to the problem may be gained by considering a simple $1D$
quantum system. We shall consider the domain walls at zero temperature of the
following extension of the spin $S$ Ising quantum chain
\begin{equation}
\label{Ham}
{\cal H} = - {H \over 2S} \sum_{n=1}^L S_x (n) - 
{J \over 2S^2} \sum_{n=0}^{L} 
S_z(n) S_z(n+1) - \frac{\eps J}{2 S^2} \sum_{n=1}^{L-1}  S_y(n) S_y(n+1)  
\end{equation}
where $S_{x,y,z}(n)$ is a quantum spin $S$ operator at site $n$, $H$ is the
transverse field, $\eps$ describes the anisotropy in the spin space, 
$J$ is the exchange coupling and $L$ is the number of sites. 
We shall work with the boundary conditions
\BEQ \label{Bound}
S_z(0) = - S_{z}(L+1) = S
\EEQ
The study of finite values of $S$ rather than considering the classical limit
$S\rar\infty$ 
may be of interest, since several experimentally studied systems have a rather
low value \cite{Mike91,Mike92} of $S$: 
$S=\frac{1}{2}$ in CsCoCl$_3$ and {\sc chab}, 
$S=1$ in CsNiF$_3$ and $S=\frac{5}{2}$ in {\sc tmmc}. Profiles of various
physical quantities have been investigated by several 
authors \cite{Mike91,Mike92,Prel81,Lima94,Henk95}. Here we shall concentrate
on the local magnetization profile
\begin{equation} \label{MagPro}
m(r) = S^{-1}\langle 0| S_z (n) |0\rangle \;\; , \;\; r = n/(L+1) \;\; , \;\;
n = 0,1,\ldots, L+1
\end{equation}
where $|0\rangle$ is the ground state of $\cal H$. In the sequel, we shall 
often use the notation $h=H/J$. 

Depending on the parameters $h$ and $\eps$, there are three distinct 
ground state phases as indicated in table~\ref{tab1}. Their properties
are distinguished \cite{Baro71} through the magnetization 
$m= \sum_r m(r)$ and the connected
two-point function $G(R) = \langle S_z(R) S_z(0)\rangle - \langle S_z(R)
\rangle\langle S_z(0)\rangle$.
\begin{table}
\caption[Phase diagram]{Ground state phase of the 
quantum Hamiltonian $\cal H$. 
The phase boundaries are for spin $S=\frac{1}{2}$. \label{tab1}}
\vspace{0.4cm}
\begin{center}
\begin{tabular}{|c|ccc|} \hline
phase & condition          & $m$      & $G(R)$ \\ \hline
D     & $\eps < h-1$       & 0        & $R^{-1/2} \exp(-R/\xi)$ \\
F     & $h-1<\eps < h^2/4$ & $\neq 0$ & $R^{-2} \exp(-R/\xi)$ \\
O     & $h^2/4 < \eps$     & $\neq 0$ & $R^{-2} \exp(-2R/\xi) \Re (B e^{iK R})$
\\ \hline
\end{tabular} \end{center}
\end{table}
The correlation length $\xi$ and the wave vector $K$ are for $S=\frac{1}{2}$ 
known\cite{Baro71} functions of $h$ and $\eps$. For higher spin 
$S>\frac{1}{2}$,
the F/O boundary is exactly known,\cite{Kurm82} while the D/F boundary 
(which is in the Ising universality class) has to
be found numerically.\cite{Hofs95} Besides studying the spin dependence of
the profile eq.~(\ref{MagPro}), we shall also ask how the 
profile changes between
the two ordered phases F and O. 

Finally, the results presented here may have a bearing on certain
non-equilibrium reaction-diffusion processes, when the master equation
is rewritten in a Hamiltonian form.\cite{Alca94} The resulting quantum
Hamiltonians are almost identical 
to eq.~(\ref{Ham}) (at the O/F boundary) and surface fields
model particle injection and extraction.\cite{Kreb95} Reactions of a single
particle species are described by spin $\frac{1}{2}$, where the free fermion 
condition means that two particles react with infinite rate on encounter.
That condition is apparently realised for the exciton kinetics 
in {\sc tmmc}.\cite{Henk95a} Ground state order parameter profiles as 
considered here should be analogous to particle density profiles in the
steady state. 

In section 2, we review the results for the order parameter profile. In 
distinction to the classical case, quantum effects lead to the delocalization
of the energy eigenstates and thus to wide profiles. 
In section 3, related spectral properties of $\cal H$ will be described. 

\section{Order parameter profile for $S=1/2,1$ and $3/2$}

\begin{figure}
\vskip 5.5cm
\psfig{figure=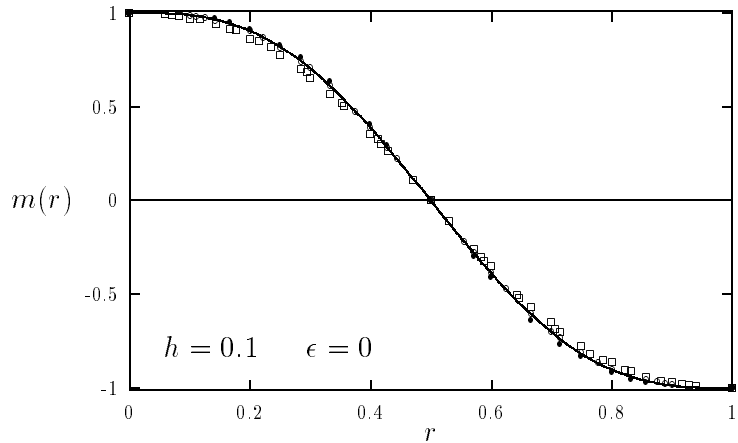,height=1.1cm}
\caption[Mag profiles]{Magnetization profiles $m(r)$ for spin $S=\frac{1}{2}$
(boxes), $S=1$ (full circles) and $S=\frac{3}{2}$ (open circles). The
curve gives the profile of eq.~(\ref{EinWall}) which is correct to leading
order in $h$. \label{fig1}}
\end{figure}

We now display
the results for the profile eq.~(\ref{MagPro}). 
On finite lattices, the calculation of the profile from the lowest eigenstate
of $\cal H$ through the Lancz\'os algorithm 
is completely standard.\cite{Chri93} We begin with the
profiles for the phase F, which is shown in Fig.~\ref{fig1}
for $S=\frac{1}{2}, 1$ and $\frac{3}{2}$.
We observe the following.\cite{Henk95} 
First, data for different lattice sizes (we used lattices up to $L=16$) nicely
collapse onto each other which implies that for the parameters considered the
lattices are large enough to faithfully represent the $L\rar\infty$ limit. 
Second, the profiles are quite wide and vary continuously with $r$. Third,
the form of the profile is, at least for $S$ not too large, quite independent
of the value of $S$. These differences grow larger when $h$ approaches the
critical value (if $\eps=0$, $h_c=1$ for $S=\frac{1}{2}$,\cite{Baro71}
$h_c \simeq 1.32587(1)$ for $S=1$\cite{Hofs95}). 
Finally, the profile only depends weakly on $h$ for $h$ small enough. 

Finding a quantum profile with a width $w \sim L$ of the order of the system
size $L$ is in contrast with the sharp profiles of width $w \sim a$, 
the lattice constant, in the classical $S\rar\infty$ limit. 
This result can be explained as follows.\cite{Mike91,Henk95} The boundary
conditions eq.~(\ref{Bound}) require that at least one domain wall is present
in the system. For $h=0$ (and $\eps=0$), 
the states of lowest energy of $\cal H$ are 
\begin{equation} \label{shalf}
\left| \uparrow \uparrow \cdots \uparrow 
\downarrow\right\rangle \;\; , \ldots , \;\; 
\left| \uparrow \cdots \uparrow \downarrow 
\cdots \downarrow \right\rangle \;\; , 
\ldots , \;\;
\left| \uparrow \downarrow \cdots \downarrow \downarrow \right\rangle
\end{equation}
for spin $S=\frac{1}{2}$ 
and for $L$ sites, there are $L+1$ of these degenerate states. 
For $S=1$ at $h=0$ the $(L+1)+L=2L+1$ states
\begin{eqnarray}
~& \left| \uparrow \uparrow \cdots \uparrow 
\downarrow\right\rangle \;\; , 
\ldots , \;\; 
\left| \uparrow \cdots \uparrow \downarrow \cdots 
\downarrow \right\rangle \;\; , 
\ldots , \;\;
\left| \uparrow \downarrow \cdots \downarrow \downarrow 
\right\rangle \nonumber \\
 & \left| \uparrow \uparrow \cdots \uparrow 0
\downarrow\right\rangle \;\; , 
\ldots , \;\; 
\left| \uparrow \cdots \uparrow 0\downarrow \cdots 
\downarrow \right\rangle \;\; , 
\ldots , \;\;
\left| \uparrow 0\downarrow \cdots \downarrow \downarrow \right\rangle
\label{sone}
\end{eqnarray}
are degenerate. These states have a single sharp domain wall. 
Turning on $h$, the 
transverse field term acts like a hopping matrix between these states and
the domain wall profile becomes broad as a result of the superposition of
many sharp domain walls. Working in the subspace $\cal M$
of the states eqs.~(\ref{shalf},\ref{sone}) 
with a single domain wall, the order parameter profile is
found to be \cite{Mike91,Henk95}
\BEQ \label{EinWall}
m(r) = 1 -2r + \frac{1}{\pi} \sin(2\pi r)
\EEQ
for both $S=\frac{1}{2}$ and $S=1$. 
As can be seen from Fig.~\ref{fig1}, for $h$ small enough eq.~(\ref{EinWall})
is in excellent agreement with the data. 

Indeed, for spin $S=\frac{1}{2}$, the correspondence
between the domain walls and the lowest excitations can be made precise
\cite{Mike91,Mike92}. Namely, for the quantum chain eq.~(\ref{Ham}) with
open boundary conditions one can define the domain wall creation 
operators, with $a_n = \frac{1}{2} \left( S_x(n)+i S_y(n) \right)$
\BEQ
D_{n}^{+} = \frac{1}{2} \left( c_{n+1}^{+} + c_{n+1} \right) 
+ \frac{1}{2} \left( c_{n}^{+} - c_{n} \right) \;\; , \;\;  
c_{n} = \exp\left[ i\pi \sum_{\ell=1}^{n-1} a_{\ell}^{+} a_{\ell} \right] a_n
\EEQ
which satisfy the anticommutation relations 
$\left\{ D_{n}^{+}, D_m \right\} = \delta_{n,m}$, 
$\left\{ D_n^{+}, D_{m}^{+} \right\} = 0$. 
$D_n^{+}$ and $D_n$, respectively,
 create and annihilate a domain wall between the sites
$n$ and $n+1$. Then, for {\it open} boundary conditions rather
than the ones specified in eq.~(\ref{Bound}) and for $S=\frac{1}{2}$, the
quantum  Hamiltonian becomes \cite{Mike91}
\BEA
{\cal H} &=& J \sum_{n=1}^{L-1} D_n^+ D_n 
+\frac{1}{2}\eps J \sum_{n=1}^{L-1} \left( 
D_{n-1}^+ D_{n+1} - D_{n-1}^+ D_{n+1}^+
+ \mbox{\rm h.c.} \right) \nonumber \\
 & & -\frac{1}{2} H \sum_{n=1}^L \left( D_{n-1}^+ D_n - D_{n-1}^+ D_{n}^+ +
\mbox{\rm h.c.} \right) -\frac{1}{2} J(L-1) \label{DomHam}
\EEA
where $D_L = D_0 = \frac{1}{2}(c_1^+ +c_1 + c_L^+ - c_L)$. The Hamiltonian
thus describes domain wall transfer between nearest neighbor and 
next-nearest neighbor sites as wall as domain wall pair creation and
annihilation. This gives a nice interpretation to the 
well-known\cite{Baro71} free fermion formulation. 
While in the F phase the next-to-nearest neighbor terms
are unimportant, this is not so in the O phase. 

\begin{figure}
\psfig{figure=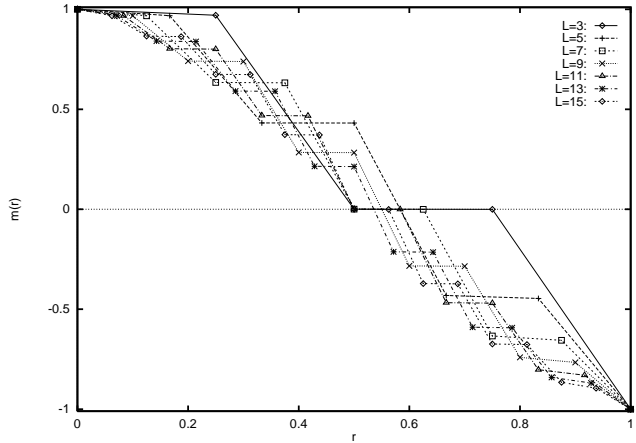,height=1.2cm}
\vskip 6cm
\caption[O-Profil]{Magnetization profile in the O phase for spin 
$S=\frac{1}{2}$ and $h=0.01, \eps=0.5$ for several values of the system
size $L$. \label{fig2}}
\end{figure}

Turning the profile in the O phase, different results are
obtained from finite lattices for $L$ even and $L$ odd.\cite{Henk95} 
For $L$ even, the
lowest ground state remains non-degenerate and eq.~(\ref{MagPro}) can be used.
That is not so for $L$ odd, where there are for $h=0$ and $\eps\neq 0$ two
degenerate ground states $\ket{0}, \ket{0'}$ and these states remain close
together even for $h\neq 0$. Then the order parameter profile is found from the
eigenvalues $m_{\pm}(r)$ of the matrix
\begin{equation} \label{OrdParMat}
\hat{m} (r) = S^{-1}  
\left( 
\begin{array}{cc} 
{\langle0|S_z(n)|0\rangle} & {\langle0'|S_z(n)|0\rangle} \\
{\langle0|S_z(n)|0'\rangle}& {\langle0'|S_z(n)|0'\rangle}
\end{array}
\right) \;\; , \;\; r=n/(L+1)
\end{equation}
and $m_+(r)$ and $m_{-}(r)=-m_{+}(\frac{1}{2}-r)$ 
are related to each other through 
spatial reflection. One of these is shown in Fig.~\ref{fig2}.
We observe the following.\cite{Henk95} First, finite-size effects are much
larger than in the F phase. Second, $m(r)$ displays for $L$ finite step-like
behavior and it looks as if the system were built from hard objects each
occupying two lattice sites. When $h$ is increased, these composites soften
until they melt at the O/F transition. Third, for $L$ odd, the reflection
symmetry is lost. These results can be explained \cite{Henk95} 
in the same way as before
through the hopping of domain walls, as described by the Hamiltonian in the
form eq.~(\ref{DomHam}). Calculations become particularly simple for $h=0$ and
in the context of the subspace $\cal M$ of the states eq.~(\ref{shalf}).
We can then decompose ${\cal M}= {\cal M}_e + {\cal M}_o$ into the subspaces
with an even and odd number of inverted spins, respectively. Then, the 
effective Hamiltonian decomposes into two which are
identical to the one found in the treatment of the
F phase, but for a redefined effective system size. 
We have $L_{\mbox{\small \rm eff}}=
\frac{1}{2}(L+1)$ for $L$ odd (twice) and 
$L_{\mbox{\small \rm eff}}=L/2 +1$ and $L/2$ for
$L$ even. It follows that in the $L\rar\infty$ limit the width of the 
terraces decreases and the order parameter profile goes toward a smooth
and reflection symmetric limit. The phenomenon seen in Fig.~\ref{fig2} is thus
a finite-size effect. 

Finally, in the $L\rar\infty$ limit, we observe (for spin $S=\frac{1}{2}$)
numerically a simple relation
of the profiles $m(r;h,\eps)$ between the F and the O phases, suggesting
that \cite{Henk95}  
\BEQ
m\left(r;h_{\mbox{\small \rm eff}}(\eps),0^{ }\right) \simeq 
m\left(r; 0^{ }, \eps\right)
\EEQ
In the limit $\eps\ll 1$, the restriction to the subspace $\cal M$ 
is accurate and yields 
$h_{\mbox{\small \rm eff}}(\eps)=\eps$. 
For $\eps$ finite, we find that phenomenologically 
$h_{\mbox{\small \rm eff}}(\eps) \simeq \sqrt{\eps}$. 

With increasing values of $S$ the mechanism of quantum delocalization through
hopping of the domain walls becomes less and less important as the classical
limit is approached.\cite{Mike91,Mike92,Henk95} Again, this is best
understood within the subspace $\cal M$ of states with a single domain 
wall. The quantum hopping of a domain wall becomes analogous to a
tunneling process, since there is at least in the semiclassical limit
$S\rar\infty$ a finite energy barrier which prevents the free motion
of the domain walls. This energy barrier vanishes for finite values of $S$
in the thermodynamic limit. 
Starting from $S=\frac{3}{2}$, the states can be such arranged that
the hopping matrix elements between them 
become periodic functions, thus leading to a
banded energy spectrum.\cite{Ashc76} The width of these bands decreases 
rapidly \cite{Mike92,Henk95} with $S$, $W \sim H 2^{-S}$. Finally, for 
$S=\infty$, quasiclassical approximations \cite{Mike92,Henk95} become accurate
and yield a sharp domain wall (for $h \ll 1$) of the order of the lattice
spacing. 

The effects of temperature and of inhomogeneous transverse fields
on the quantum profile have also been studied.\cite{Lima94} The constant
transverse field $H$ is replaced by $H(n)$ which takes the value $+H$ in the
left sector and $-H$ in the right sector of the system. 
Then for $S=\frac{1}{2}$
and $\eps=1$, the profile $e(n)=\bra{0} S_x(n)\ket{0}$ can be calculated
exactly. In distinction to the monotonous profiles of 
Figs.~\ref{fig1},\ref{fig2}, $e(n)$ is found to show oscillatory behavior in
the O phase. For $T\rar 0$, these oscillations are suppressed. 

\section{Nature of the spectrum}

We have seen that at least qualitatively, the form of the profile $m(r)$
can be understood in terms of the subspace $\cal M$ of the states with a single
domain wall. While this approximation is {\it a priori} applicable for
$h\ll 1$,it is remarkable that the clear separation of the states within
$\cal M$ remains intact even for finite values of $h$.\cite{Henk95}
We can thus identify
a subspace $\wit{\cal M}$ of low-lying states (with ${\cal M} = \lim_{h\rar 0}
{\wit{\cal M}}$) even though the states in $\wit{\cal M}$ are
no longer simply those states with a single domain wall. Indeed, we find that
while the energy gaps of the states outside $\wit{\cal M}$ 
are {\it finite} within
the phases O and F, the states within $\wit{\cal M}$ show a different behavior.

We begin with the F phase. Let $g(i) = E_i - E_0$ denote the i$^{th}$ gap.
We then find \cite{Henk95} the finite-size 
scaling $g(i) \sim L^{-\theta}$ with
$\theta=2$ for the $2SL+1$ states within $\wit{\cal M}$. 
Furthermore, the scaling
amplitudes $a(i) = \lim_{L\rar\infty} L^2 g(i)$ follow a simple pattern
within $\wit{\cal M}$. Namely ($i=1,2,3\ldots$) 
\BEQ \label{AmpF}
r(i) = a(i)/a(1) = \frac{1}{3}i(i+2) = 1,\frac{8}{3},5,8,\frac{35}{3},16,\ldots
\EEQ
We have checked these amplitude ratios within the entire F phase for both
$S=\frac{1}{2}$ and $S=1$. This confirms the earlier observation \cite{Mike92}
that domain walls approach their asymptotic behavior algebraically. 

\begin{figure}
\psfig{figure=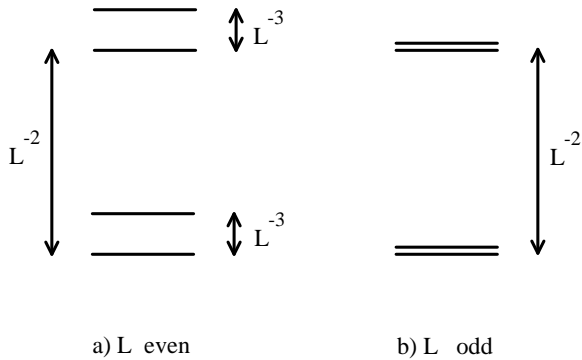,height=3.3cm}
\caption[Lacunes]{Schematic structure of the low-lying states of $\cal H$ for
$S=\frac{1}{2}, h=0$. \label{fig3}}
\end{figure}

For the O phase, the spectrum is different. We concentrate on $S=\frac{1}{2}$
and $h=0$. The basic structure of the levels is indicated in 
Fig.~\ref{fig3}.\cite{Henk95} 
For $L$ even, the states within $\wit{\cal M}$ 
are grouped into doublets with a
splitting of order $L^{-3}$ and the gaps between the doublets are of order
$L^{-2}$. Thus the lowest gap scales with an exponent $\theta=3$, while
$\theta=2$ for all other gaps. 
For $L$ odd, all levels are doubly degenerate with gaps between them
of order $L^{-2}$ and $\theta=2$ throughout. For the amplitudes, we find 
independently of the parity of $L$ the
following pattern, consistent with the 
structure \footnote{Eqs.~(\ref{AmpF},\ref{AmpO})
can be analytically reproduced by restricting to the space $\cal M$.} 
eq.~(\ref{AmpF})
\BEQ \label{AmpO}
\rho(i) = a(i)/a(2) = 0, 1, 1, \frac{8}{3}, \frac{8}{3}, 5, \ldots
\EEQ
along the entire line $h=0$. Qualitatively, the same structure persists through
the entire O phase.

\section*{Acknowledgements}
It is a pleasure to thank A.B. Harris and M. Cieplak for the enjoyable 
collaboration which has led to the work reviewed here.

\end{document}